\documentclass[article]{revtex4-1} 
\usepackage{graphicx}
\usepackage{subfig}
\usepackage{color}
\usepackage{epstopdf}

\begin{document}
\title{ Natural focusing of symmetric Airy beams}
\author{R. J\'auregui}
\affiliation{Instituto de F\'{\i}sica, Universidad Nacional Aut\'onoma de M\'exico}
\address{Apartado Postal 20-364, 01000, M\'exico D.F., M\'exico.}
\email{rocio@fisica.unam.mx}
\author{P. A. Quinto-Su}
\affiliation{Instituto de Ciencias Nucleares, Universidad Nacional Aut\'onoma de M\'exico}
\address{Apartado Postal 70-543, 04510, M\'exico D.F., M\'exico.}

\begin{abstract}
In this work we study the natural focusing of Airy beams symmetric under reflection of the transverse coordinates. Following a recent proposal, their angular spectra depend on the absolute value of the third power of the transverse components of the wave vector.  We show that these beams are related to Airy and Scorer special functions.
The caustics have a morphology that does not correspond to the ones described by standard catastrophe optics. The structural stability of symmetric Airy beams is experimentally probed.
\end{abstract}


\maketitle

Several wave phenomena give rise to bright focusing features formed by the 
coalescence of multiple rays, {\it i.e.}, caustics. These features are found in  both classical and quantum  scenarios. The usefulness of bright focusing is vast and diverse including gravitational microlensing enhancement \cite{grav}, manipulation of electron waves \cite{KAidala} and laser light shaping \cite{dholakia-shaping}. At the end of the 1970's, it was  found that the geometry, scaling properties and diffraction patterns derived from natural focusing can be described in terms of elementary catastrophe classes \cite{berry1989,nye}. In optics, the simplest class realization corresponds to Airy beams which have many interesting properties that have been exploited in a wide range of applications \cite{kasparian,shang,dolev}.
Extending the concept of Airy beams by making them symmetric under reflection of the transverse coordinates changes some of their  basic properties. The experimental generation of these beams \cite{brasil} and a first study of their usefulness for optical trapping \cite{optrapping} have been  reported recently. Here we show, that these symmetric beams can be mathematically related to a complex superposition of
Airy (Ai) and Scorer (Gi)  special functions, hence in the following we name them Airy-Scorer beams. They  naturally focus  in free space
yielding caustics with a nontrivial morphology that does not correspond to those described by standard catastrophe optics.  We also report the experimental probe of the structural stability of Airy-Scorer beams. 

Airy optical beams \cite{siviloglou} are physical realizations of wave packet solutions
of the paraxial equation with an angular spectra dependent
on the third power of the transverse component of the wave vector:
\begin{eqnarray}
\phi_a^{(1)}(s,\zeta)&=&\frac{1}{2\pi}\int_{-\infty}^\infty dk_s \mathfrak{F}^{(1)}_a(k_s)e^{i(k_s s  -\frac{k_s^2\zeta}{2})}\nonumber\\
 \mathfrak{F}^{(1)}_a(k_s) &=& e^{(a-ik_s)^3/3}.\label{ai:fs}
\end{eqnarray}
Here, $s =x/x_0$ ($k_s$) is a dimensionless transverse coordinate (wave vector) component, $x_0$ is a length scale, $\zeta = z/k x_0^2$ is
the propagation coordinate measured with respect to the Rayleigh scale, and $k =2\pi n / \lambda_0$ is
the wave number of the monochromatic optical wave.

 Airy optical beams exhibit remarkable properties
predicted by Berry and Balazs \cite{berry1979} taking as starting point their ideal mathematical representations,
in terms of  the Airy Ai functions:
\begin{eqnarray}
\phi_a^{(1)}(s,\zeta) &=& \mathrm {Ai}(s-\zeta^2/4 + ia\zeta) e^{as - a\zeta^2/2 -i\varphi},\label{eq:aifp}\nonumber\\
\varphi &=&\zeta^3/12 - a^2\zeta/2 -s\zeta/2,
\end{eqnarray}
for the case $a$=$0$ \cite{footnote}.
Three dimensional Airy beams result from considering the two dimensional angular spectrum
\begin{equation}
\mathfrak{F}^{(2)}_{a_1,a_2}(k_{s_1},k_{s_2}) = e^{(a_1 + ik_{s_1})^3/3}e^{(a_2 + ik_{s_2})^3/3}.\label{ai:fs2}
\end{equation}
 Their experimental implementation \cite{siviloglou} is  relatively simple; it requires a Gaussian beam modulated by a cubic phase and a lens to perform the proper Fourier transform.

 Airy beams remain diffraction-free over long distances while they tend to freely accelerate during propagation.
This  can be understood by considering the curves at which the real part of the argument of the Airy function 
 in $\phi_a^{(1)}(s,\zeta)$ becomes constant, which are parabolas in $(s,\zeta)$ space.
Another interesting property of Airy-beams is that they are self-healing: they tend to reform during propagation in spite of blocking a part of them. This property is behind their robustness in scattering and turbulent environments \cite{broky}.

\begin{figure}[ht]
\centering\includegraphics[width=8cm]{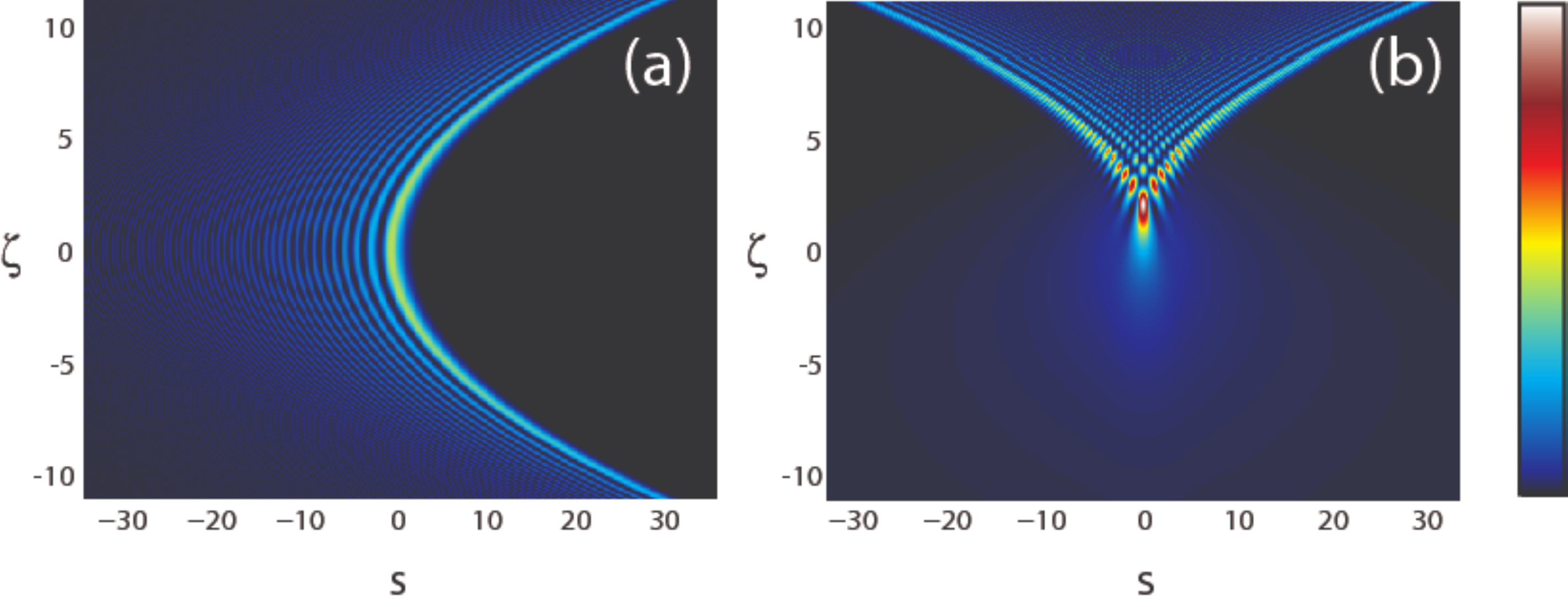}
\caption{(a) Intensity pattern of a two dimensional (a)Airy beam, $\vert\phi^{(1)}_a(s,\zeta)\vert ^2$; (b)Airy-Scorer beam
 $\vert\Phi_{a}^{(1)}(s,\zeta)\vert^2$. The observed structures are caustics, $i.$ $e.$  surfaces arising from natural focusing. The parameter $a$ is taken as 0.01 in both figures and the maximum intensity in (b) equals more than two  times the maximum intensity in (a).}\label{fig:1}
\end{figure}

 From a  geometrical optics point of view, Airy beams illustrate paradigmatic
examples of optical catastrophes \cite{berry1989,nye,Vo2010,Yan2012}.
Two dimensional Airy beams exhibit  a fold diffraction catastrophe \cite{berry1979}.
For $s<0$ there are two stationary points in the exponent of the integrand in Eq.~(\ref{ai:fs}) whose
interference gives rise to oscillations in $\phi_0^{(1)}(s,0)$,
while for $s>0$, there is only one stationary point which in fact lies in the imaginary axis,
leading to an exponential decay towards zero of $\phi_0^{(1)}(-s,0)$. Also notice
that $a>0$ gives rise to a localization of the Airy beam along its main propagation axis, that is
$\vert\phi_a^{(1)}(s,\zeta)\vert^2 <<1$ as $\vert\zeta\vert>>1$. In Fig.~\ref{fig:1}, the intensity
 of a two dimensional Airy beam, $\vert\phi_a^{(1)}(s,\zeta)\vert ^2$,
 is illustrated.

\begin{figure}[hb]
\centering
\includegraphics[width=8cm]{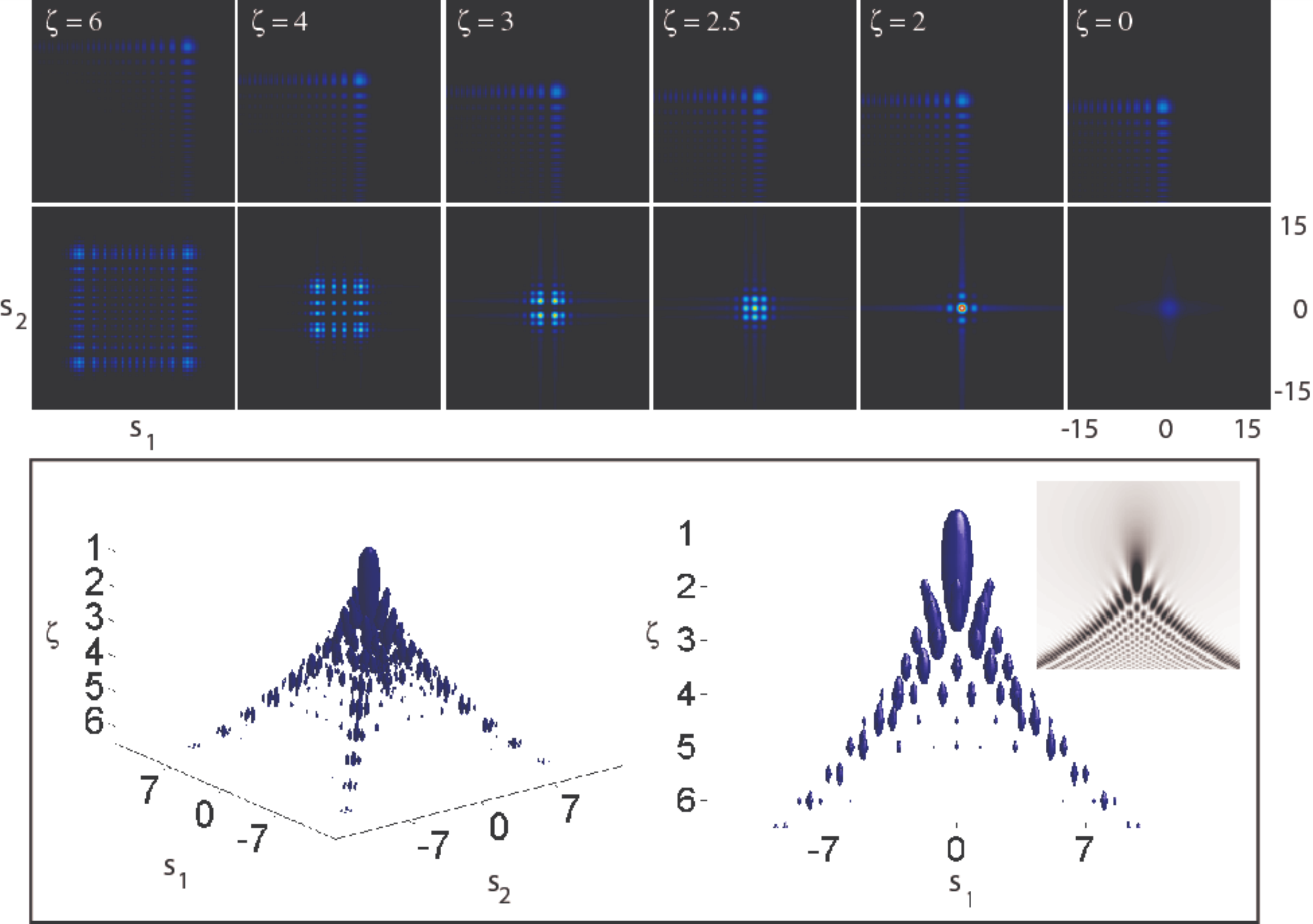}
\caption{In the first row the intensity pattern of a three dimensional Airy beam $\vert\phi^{(2)}_a(s,\zeta)\vert ^2$ is illustrated at different sections with constant $\zeta$ values. In the second row the intensity pattern of a three dimensional Airy-Scorer beam  $\vert\Phi_{a}^{(2)}(s,\zeta)\vert^2$ is illustrated for similar values of $\zeta$. In the third row a  3d reconstruction of the surfaces with intensities is shown from different view angles. The small figure at the right corner in the third row shows the corresponding caustics profile for the two dimensional Airy-Scorer beam. In all the shown figures the parameter $a=0.01$.}\label{fig:2}
\end{figure}

 In Ref.~\cite{Vo2010}, the structure of three dimensional Airy beams is interpreted as an aberrated wavefront
and the lateral shift of Airy beams during propagation is presented in the context of the
three dimensional caustic representation \cite{ft1}. In Fig.~\ref{fig:2}
the intensity of a three dimensional Airy beam, $\vert\phi_a^{(2)}(s_1,s_2,\zeta)\vert ^2$, is exemplified.

Here we study a direct generalization of Airy beams symmetric
under reflections of the transverse coordinates $s_i\rightarrow-s_i$. There is not an unique way to impose such condition.
 In this work we set a Fourier angular spectra symmetric under reflection of the transverse wave vector coordinate 
 $k_{s_i}$. In the following, we study their general properties including their comparison to Airy beams.

The Fourier spectra  of two dimensional Airy-Scorer beams is taken as
\begin{equation}
\mathfrak{S}^{(1)}_a(k_s) = e^{(a -i\vert k_s\vert )^3/3}.
\end{equation}
The relation to Airy like functions is  found by considering the resulting wave packet
\begin{equation}
\Phi_{a}^{(1)}(s,\zeta) = \frac{1}{2\pi}
\int_{-\infty}^\infty dk_s \mathfrak{S}^{(1)}_a(k_s)e^{i(k_s s  -\frac{k_s^2\zeta}{2})}\label{eq:ai.gi}
\end{equation}
evaluated at $\zeta = 0$ and for $a=0$:
\begin{eqnarray}
\Phi_{0}^{(1)}(s,0) &=& \frac{1}{2\pi}\int_{-\infty}^\infty dk_s \mathfrak{S}^{(1)}_0(k_s)e^{ik_s s}\\
&=&\frac{1}{2\pi}\int_{0}^\infty dk_s\Big(e^{i(k_s^3/3 + k_s s)} +e^{i(k_s^3/3 - k_s s))}\Big)\nonumber\\
&=& \frac{1}{2}\Big[(\mathrm{Ai}(s)+i\mathrm{Gi}(s)) +(\mathrm{Ai}(-s)+i\mathrm{Gi}(-s))\Big]\nonumber,
\end{eqnarray}
with $\mathrm{Gi}$ a Scorer function \cite{nist} which has the integral representation
\begin{equation}
\mathrm{Gi}(\eta) =\frac{1}{\pi}\int_0^\infty \sin(\frac{1}{3}t^3 + \eta t)dt;
\end{equation}
 complementing  the  Airy Ai functions integral representation
\begin{equation}
\mathrm{Ai}(\eta) =\frac{1}{\pi}\int_0^\infty \cos(\frac{1}{3}t^3 + \eta t)dt.
\end{equation}

The function $\mathrm{Gi}(\eta)$ satisfies the inhomogeneous  differential equation
\begin{equation}
\mathrm{Gi}^{\prime\prime} + \eta\mathrm{Gi} =-\frac{1}{\pi}.
\end{equation}
It is related to the Airy $\mathrm{Bi}$ functions by the relation
\begin{equation}
\mathrm{Gi}(\eta) = \mathrm{Bi}(\eta) - \frac{1}{\pi}\int_0^\infty e^{-\frac{1}{3}t^3 + \eta t} dt
\end{equation}
so that, contrary to the Airy $\mathrm{Bi}$ function, $\mathrm{Gi}\rightarrow 0$ as $\eta\rightarrow\infty$.
The inhomogeneous term in the differential equation for Gi prevents the existence of
a closed analytical expression analogous to Eq.~(\ref{eq:aifp}) for the propagated beam $\Phi_{a}^{(1)}(s,\zeta)$ in terms of the Ai and Gi functions.

Two dimensional Airy-Scorer (Ai-Gi) beams do not exhibit a fold diffraction catastrophe. For $\zeta = 0$ and any given value of $s$, there is just one stationary point in the exponent of the integrand. However these beams do exhibit  interesting caustics as clearly exemplified in Fig.~\ref{fig:1}. The stationary points of the exponent in the integrand  defining $\Phi_{0}^{(1)}(s,\zeta)$, Eq.~(\ref{eq:ai.gi}), on the section $s=0$ are  two along the positive values of the propagation direction coordinate,  $\zeta>0$. The accompanying  interference gives rise to oscillations in $\Phi_{0}^{(1)}(0,\zeta)$. For $\zeta<0$, there is only one stationary point which in fact lies in the imaginary axis, leading to an exponential decay towards zero of $\Phi_{0}^{(1)}(s,0)$ as a function of $\zeta$.
The cusp diffraction catastrophe characterized by the germ $k_s^4/4 + k_s \zeta^2/2 + k_s s$ has a quite similar structure to
the one obtained here \cite{berry1989}. The main reason is that the absolute vale of $k_s$ has the same parity than $k_s^4$. Notice, however, that the difference in the power of $k_s$ gives rise to different scaling behaviors between the standard cusp catastrophe and that arising in Ai-Gi beams. Besides two dimensional Ai-Gi beams do not have the wave front dislocations typical of the beams derived from a $k_s^4/4 + k_s \zeta^2/2 + k_s s$ germ.

The Fourier spectra of three dimensional Ai-Gi beams
\begin{equation}
\mathfrak{S}^{(2)}_{a_1,a_2}(k_{s_1},k_{s_2}) = e^{(a_1 -i\vert k_{s_2}\vert )^3/3}e^{(a_2 -i\vert k_{s_1}\vert )^3/3}
\end{equation}
give rise to  the following wave packets
\begin{eqnarray}
\Phi_{a_1,a_2}^{(2)}(s_1,s_2,\zeta) &=& \frac{1}{(2\pi)^2}
\int_{-\infty}^\infty dk_{s_1}dk_{s_2} \mathfrak{S}^{(2)}_{a_1,a_2}(k_{s_1},k_{s_2})\nonumber\\&\cdot&
e^{i(k_{s_1} s_1  -k_{s_1}^2\zeta/2)}e^{i(k_{s_2} s_2  -k_{s_2}^2\zeta/2)}.\label{eq:Phi2}
\end{eqnarray}
The morphology of these beams inherits many properties from its 2d analogue.
In Figs.~\ref{fig:2} the  intensity of a three dimensional Ai-Gi beam, $\vert\Phi_a^{(2)}(s,\zeta)\vert ^2$, is illustrated. While for an standard three dimensional Airy beam the diffraction catastrophe is qualified as a hyperbolic
umbilic\cite{Vo2010,Yan2012}, the optical caustics of a three dimensional Ai-Gi beam
consist of a tightly localized high intensity spot surrounded by fringes receding
to four equidistant asymptotes $2\pi/4$ apart in a pyramid like structure. The general morphology of the caustics resembles the superposition of four Airy beams with the main lobes positioned at the vertices of a square with the tails pointing toward the center. Notice that the general properties of the optics catastrophe  in Ai-Gi beams do not correspond to any of the standard
classifications in terms of germ polynomials due to the presence of an absolute value in the cubic terms.

The numerical simulations predict a high natural focusing arising from imposing that the beam phase be consistent with the symmetric complex superposition of Airy and Scorer functions in the $\zeta = 0$ plane . This three dimensional beam focusing yields well localized light droplets for $\zeta>0$ and a rapidly vanishing intensity for the region $\zeta<0$.
Under similar conditions, the maximum in intensity of a two dimensional Ai-Gi beam doubles that of the corresponding Airy beam; for the three dimensional situation the maximum  of $\vert\Phi_{a_1,a_2}^{(2)}(s_1,s_2,\zeta)\vert^2$ is almost five times
 its Airy analog.

\begin{figure}[htb]
\centering\includegraphics[width=9cm]{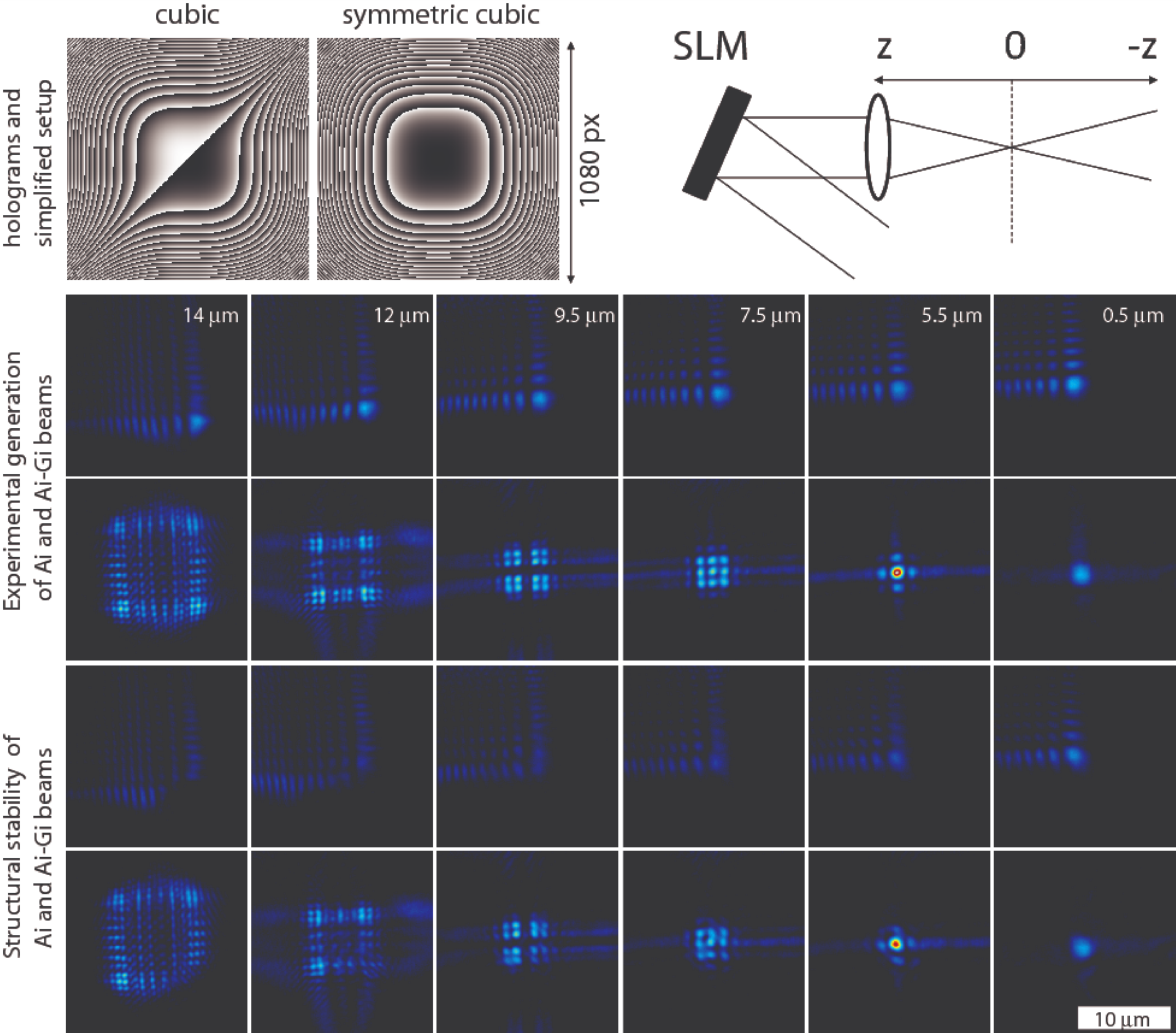}
\caption{The first row depicts the digital holograms used to generate Ai and Ai-Gi
beams, and a simplified experimental setup.
Photographs of the generated beams (Ai and Ai-Gi) in different planes
along the optical axis are shown in the next two rows.
Bottom rows show self healing in the case of the Ai beam (blocking main
lobe) and structural stability for the Ai-Gi beam when one of
the corners (lower-right) is blocked. }\label{fig:3}
\end{figure}

Following a standard procedure and similarly to Ref.\cite{brasil}, Airy and Airy-Scorer beams have been generated using a programable phase plate or spatial light modulator (SLM). An expanded
Gaussian laser beam ($\lambda =975~\mu$m) was reflected from a SLM that imprints a cubic and a symmetric
cubic phase respectively. The digital holograms were calculated with the expressions: $f_{Ai}(x, y)= (x/x_0)^3+
(y/y_0)^3$ and $f_{Ai-Gi}(x, y)= \vert x/x_0\vert^3+ \vert y/y_0\vert^3$, where
the coordinates are the coordinates of the pixels, that have a size of
$8~\mu$m and the parameter $x_0=y_0=1$ $\mathrm{mm}$. Images of the
the digital holograms are shown at the top of Fig.~\ref{fig:3}, along with a
simplified experimental setup, that shows the SLM and a Fourier lens.
The optical axis is drawn according to Eq.~(\ref{eq:Phi2}) with the
origin  set at the geometrical focus of the lens
 with positive distances before the Fourier plane.
In the experimental setup, relay optics image the plane of the
SLM at the back aperture of a microscope objective (100x, 1.25 NA)
just like in standard holographic tweezers experiments \cite{dholakia,optrapping}.
The transverse size of the beam reflected at the SLM is reduced by a
factor of two by the relay lenses.
The beam emerging from the objective is back reflected by the front face
of a glass coverslide mounted in a 3d piezo translating stage
that moves in the direction of the optical axis in steps of half a micron.
The retroreflected light is imaged by a CCD for the
different transverse planes in the range between $14~\mu$m and $0.5~\mu$m
(zero is located at the geometrical focus of the microscope objective).

The first two rows of photographs in Fig.\ref{fig:3} show the transverse planes
of Ai and Ai-Gi  beams for $z$ in the
range of $14-0.5~\mu$m.
In the first frame for both cases (far left) it can be seen that the Ai
beam essentially makes the lower left  corner of the Ai-Gi,
although the corners of the Ai-Gi show interference due to the
superposition of the the principal lobe of an Ai beam and the tails of
the main lobes located at the remaining vertices.
The intensities of the localized spots in the Ai-Gi case are much larger
than the intensity of the main lobe of the Ai
beam, which is also consistent with the simulations.

Next, we probe the structural stability of the Ai-Gi beam by blocking the
lower right corner, that is, the main lobe of the Ai like beam at $z=14~\mu$m.
Photographs of the transverse planes for the Ai and the Ai-Gi beams
are in the bottom two rows in Fig.~3.
As expected, the Ai beam recovers the main lobe (self healing) further
down the optical path.
The Ai-Gi beam is not adiffractional, however, the morphology of the transverse
planes is essentially recovered after blocking one of the corners.
Hence the self-healing property of the Ai lobes has an analog
for the Ai-Gi beam.

In this article Airy-Scorer beams have been studied and compared to standard Airy beams. It has been found
that the Airy-Scorer electromagnetic field is highly focused; the morphology of the caustics does not correspond to the generic classification of catastrophe theory due to the forced symmetry imposed by the absolute value. The main features of the caustics structure are now observed along the propagation direction. The structural stability allows both their generation by imposing the adequate boundary conditions in a spatial light modulator and  a morphology of the transverse
planes that remains essentially the same after blocking part of the beam. The increased gradients of the intensity of the  Airy-Scorer with respect to the Airy beams make interesting to study its use for optical trapping of atoms, nanostructures and microparticles \cite{optrapping}.

{\bf Acknowledgments} We thank  R. Guti\'errez-J\'auregui  for useful discussions and R. Guti\'errez-Cuevas
for performing exploratory experiments. 


\begin{thebibliography}{99}
\bibitem{footnote} The parameter $a>0$ was introduced in Ref.~\cite{siviloglou} to
provide a scenario closer to optical realizations. It gives rise to an effective cut off on the wave vector component $k_s$; let us recall that
the paraxial equation is a good approximation to the wave equation only if
$k_s\ll kx_0$.
 Other authors replace the usage of the parameter $a$ by the assumption of a
Fourier spectrum  limited by an aperture within which the amplitude is uniform, i. e.,
$\mathfrak{F}_a(k_s)\rightarrow \mathfrak{F}_{a=0}(k_s)\Theta(k_{max}-\vert k_s\vert)$
with $\Theta(x)$ the step function.
\bibitem{ft1}To that end, it results convenient to replace
$\mathfrak{F}^{(2)}_{a_1,a_2}(k_{s_1},k_{s_2})\rightarrow \mathfrak{F}^{(2)}_{0,0}(k_{s_1},k_{s_2})\Theta(k_{max}-\sqrt{k_{s_1}^2 + k_{s_2}^2}$).
\end{thebibliography}
\end{document}